\PassOptionsToPackage{colorlinks=true,citecolor=blue,linkcolor=blue}{hyperref}
\documentclass{aa}

\usepackage{hyperref}
\usepackage{keyval}
\usepackage{natbib}
\usepackage{graphicx}
\usepackage{txfonts}
\usepackage{lipsum}
\usepackage{subcaption}         
\usepackage{placeins}           
\begin{document}

   \title{Bayesian imaging inverse problem with scattering transform}



   \author{Sébastien Pierre\inst{1}, Erwan Allys\inst{1}, Pablo Richard\inst{1},  Roman Soletskyi\inst{1}, \and Alexandros Tsouros\inst{1}
   }
   \authorrunning{S. Pierre et al.} 

   \institute{Laboratoire de Physique de l’École Normale Supérieure, ENS, Université PSL, CNRS, Sorbonne Université, Université Paris Cité,
75005 Paris, France.
}

   \date{\today
   \vspace{-0.4cm}}


  \abstract
{Bayesian imaging inverse problems in astrophysics and cosmology remain challenging, particularly in low-data regimes, due to complex forward operators and the frequent lack of well-motivated priors for non-Gaussian signals.
In this paper, we introduce a Bayesian approach that addresses these difficulties by relying on a low-dimensional representation of physical fields built from Scattering Transform statistics. This representation enables inference to be performed in a compact model space, where we recover a posterior distribution over signal models that are consistent with the observed data. We propose an iterative adaptive algorithm to efficiently approximate this posterior distribution.
We apply our method to a large-scale structure column density field from the \textit{Quijote} simulations, using a realistic instrumental forward operator. We demonstrate both accurate statistical inference and deterministic signal reconstruction from a single contaminated image, without relying on any external prior distribution for the field of interest.
These results demonstrate that Scattering Transform statistics provide an effective representation for solving complex imaging inverse problems in challenging low-data regimes. Our approach opens the way to new applications for non-Gaussian astrophysical and cosmological signals for which little or no prior modeling is available.
\vspace{0.2cm}}

   \keywords{Methods: statistical - Methods: data analysis - Techniques: image processing - Large-scale structure of Universe 
   \vspace{-0.5cm}}

   \maketitle
   \nolinenumbers

\section{\label{sec:introduction}Introduction}

\begin{figure*}[h!]
\includegraphics[scale=0.35]{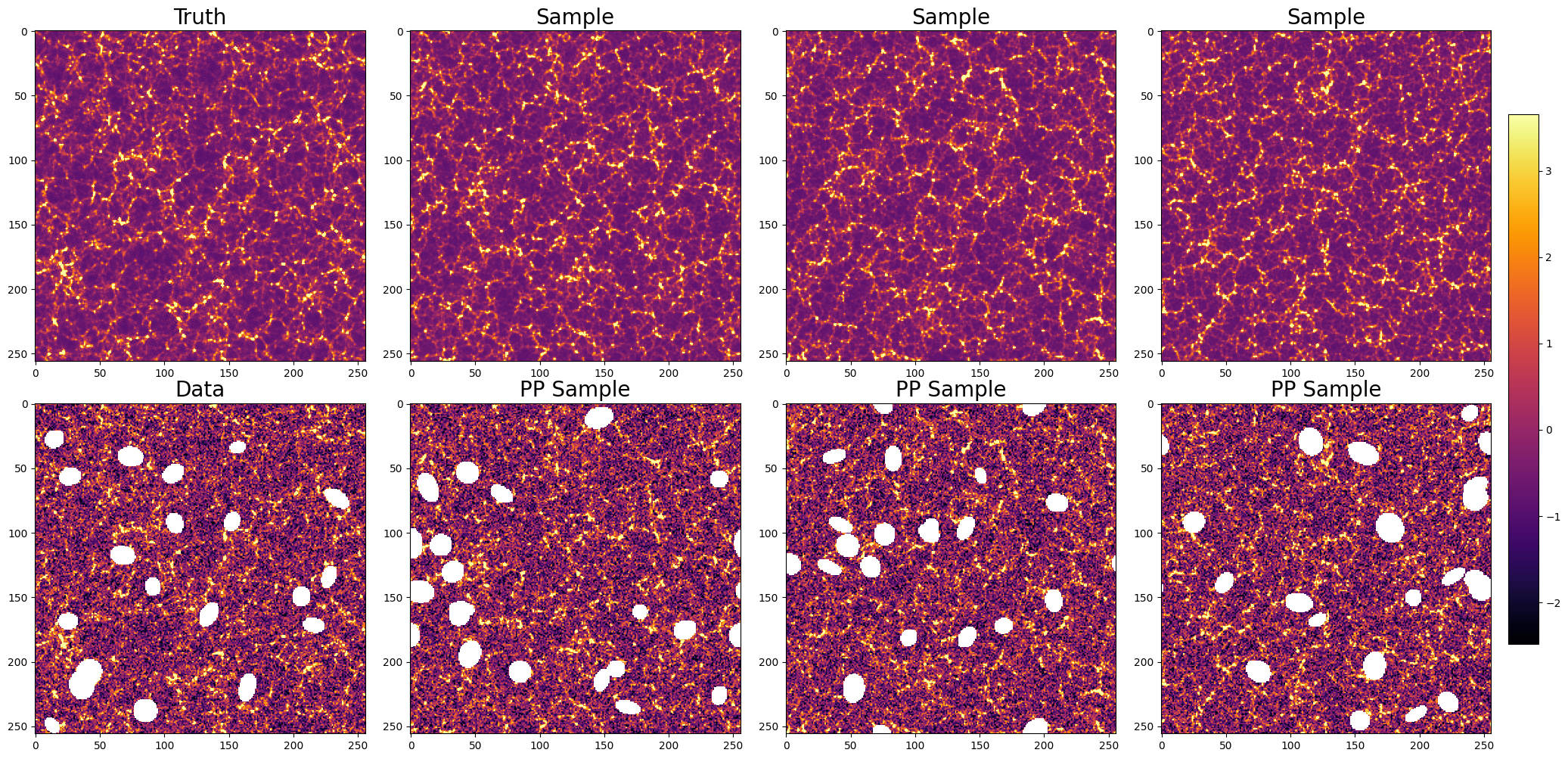}
\caption{\textbf{Top:} True field $s_{0}$ and three fields generated from ST statistics sampled from the posterior $p(\mu_{S} \mid \phi(d_{0}))$. 
\textbf{Bottom:} Observation $d_{0}$ and the corresponding posterior predictive samples obtained by applying the pixel-space forward operator $F$ to the generated fields. 
The predictive samples are visually indistinguishable from the observation, while the generated fields seem consistent with LSS maps, with variability primarily at the smallest, noise-dominated scales.}
\label{fig:visual samples}
\end{figure*}

Imaging inverse problems \citep{Bertero1998} are ubiquitous in astrophysics and cosmology. They can be formulated as the recovery of an underlying field $s_{0} \in \mathbb{R}^{n}$ from observed data $d_{0} \in \mathbb{R}^{n'}$ produced by a known forward operator $F$,
\begin{equation}
    d_{0} = F(s_{0}).
\end{equation}
This general formulation contains a wide range of astrophysical problems, including for instance weak lensing mass-mapping \citep{Kaiser1993,Remy2023}, the reconstruction of the initial density fluctuations of the universe from large scale structures \citep{Legin2023},  component separation of Galactic and cosmological signals \citep{Planck2008,Auclair2024}, the reconstruction of the Galactic Faraday rotation sky \citep{Hutschenreuter2022}, the mapping of the local Galactic dust distribution \citep{Edenhofer2024}, and the reconstruction of the local Galactic magnetic field structure from polarization data \citep{Tsouros2024}. 

Because $F$ is often stochastic or non-invertible, many candidate signals $s$ may be consistent with the same $d_{0}$. The task is then not to find a single solution, but to characterize the posterior distribution $p(s \mid d_{0})$ of signals compatible with the data. To solve such inverse problems, a Bayesian setting is typically introduced in which existing knowledge about the field of interest is encoded in a prior distribution $p(s)$. In this setting, the likelihood $p(d \mid s)$ quantifies how probable the observed data is given a candidate signal. Once the forward operator $F$ is specified it completely determines the probability distribution of any observation $d$ conditional on a given signal $s$, and therefore 
the likelihood $p(d_{0} \mid s)$. The posterior distribution is then given by:
\begin{equation}
    p(s \mid d_{0}) \propto p(d_{0} \mid s)\,p(s).
\end{equation}

This formulation allows to incorporate previous knowledge and quantify uncertainties in the recovered signal. However, specifying a physically motivated prior $p(s)$ encoding non-Gaussian information for complex processes is challenging. Data-driven priors are also limited when only few observations
are available. In this paper, we address the challenging problem of imaging inverse problems for non-Gaussian regular fields in the absence of physically motivated priors and under a very low-data regime, \textit{i.e.}, when only a single image, $d_0$, is observed.

To address the lack of physically driven models, we propose an approach based on scattering transform (ST) statistics~\citet{Mallat2012,Allys2019,Cheng2024}. These statistics $\phi(s) \in \mathbb{R}^{l}$ are designed to study non-Gaussian processes by characterizing the interaction between their different oriented scales. One major advantage of these statistics is that they can be used to construct very efficient maximum entropy generative models $p_{\mu_{S}}(s)$ of physical fields \citep{Bruna2018}, which are parametrized by the ST statistics $\mu_{S}$ themselves. In this work, the ST statistics are written $\mu_{S}$ when they parametrize maximum entropy generative models, and $\phi(s)$ when they are descriptive statistics. Sampling these gradient descent generative models, which can be compressed to a dimension $l = \mathcal{O}(10^{2})$~\citep{Cheng2024}, reproduce many regular non-Gaussian fields both visually and statistically~\citep{Allys2020,Mousset2024}. Importantly, these statistics can be robustly estimated from only a small number of samples, making them well suited to the low-data regime~\citep{Richard2025}.

In this paper, we shift the inference objective, in a Bayesian setting, from the posterior over fields $p(s \mid d_{0})$, in the high-dimensional pixel space $n = \mathcal{O}(10^{5})$, to the posterior over ST statistics $p(\mu_{S} \mid \phi(d_{0}))$, in a space of much lower dimension $l = \mathcal{O}(10^{2})$. This dimensionality reduction makes the inference problem computationally more tractable by encoding the essential non-Gaussian properties of the field in a lower dimensional manifold. This boils down to inferring a ST-based generative model of the signal $s$ rather than its exact realization $s_{0}$. Once a posterior distribution over models $p(\mu_{S} \mid \phi(d_{0}))$ is obtained, the corresponding generative models can be sampled, yielding maps from which any other statistic may be estimated. These models can also be used to train a neural network to recover a pixel-level estimate of $s_{0}$.

In this work, we propose and numerically validate an algorithm for solving complex imaging inverse problems in a single observation data regime, targeting non-Gaussian, regular fields for which no external physical prior or realistic training dataset is available. The algorithm yields an approximate posterior distribution of models $p(\mu_{S} \mid \phi(d_{0}))$ for the underlying field $s$. We present a proof of concept using large-scale structure maps from the \textit{Quijote} simulations~\citep{VillaescusaNavarro2020} and a mock instrumental forward contamination model consisting of white Gaussian noise and masks (see Fig.~\ref{fig:visual samples}). We validate that our approach allows both a statistical and deterministic reconstruction.
Although demonstrated on this specific example, the approach is general and may be applied to other cases.

This paper is organized as follows. Section~\ref{sec:Problem formulation and algorithm} introduces the problem formulation and the algorithm used to approximate the posterior. Section~\ref{sec:results} describes the data and forward operators $F$ used for validation as well as the results. We conclude this paper in Section~\ref{sec:conclusion}. Additional figures and mathematical details are provided in Appendix~\ref{app:appendices}.
\vspace{-0.3cm}
\section{Problem formulation and algorithm}\label{sec:Problem formulation and algorithm}

\subsection{Problem Formulation}
\label{sec:Formulation problem }

We consider the inverse problem $d_{0} = F(s_{0})$,
where the forward operator $F$ is known, $d_{0}$ the single observational data, and $s_{0}$ the signal we aim at constraining, that we assumed sampled from a process $S$. In our approach, the goal is to infer a ST generative model  $p_{\mu_S}(s)$ of $S$, parametrized by the ST statistics $\mu_{S} \in \mathbb{R}^{l}$, given the ST statistics of an observation $\phi(d_{0}) \in \mathbb{R}^{l}$. In a Bayesian setting, this can be done by defining a prior distribution $p(\mu_{S})$ over ST statistics, and estimating the likelihood distribution $p(\phi(d) \mid \mu_{S})$. One can then evaluate the posterior distribution conditioned on the observation $\phi(d_{0})$: 
\begin{equation}
\label{EqPosteriorDef}
    p(\mu_{S} \mid \phi(d_{0})) \propto p(\phi(d_{0}) \mid \mu_{S})p(\mu_{S}).
\end{equation}
For this project, we opted for a uniform prior distribution $p(\mu_S)$ across the entire $\mu_{S}$ space, due to the absence of a straightforward method to incorporate additional existing knowledge without introducing bias. All prior information about $s$ is then contained in its description by an ST-based generative model, which introduces in practice a very strong $p(s)$ implicit prior.

Our approach then boils down to estimating the likelihood distribution
$p(\phi(d)\mid\mu_{S})$. In standard pixel-space inverse problems, the
likelihood $p(s \mid d)$ is usually defined through the forward operator $F$, which maps any signal realization $s$ to data $d$. In contrast, our inference is performed in the ST statistics space. This requires defining a forward operator $\mathcal{F}$ in ST space, from the $\mu_S$ parameter space to the $\phi(d)$ data space:
\begin{equation}
\label{eq:effective forward}
    \mathcal{F}\hspace{-0.08cm}: ~~ \mu_{S} \;\longrightarrow\; p_{\mu_{S}}(s) \;\longrightarrow\; s 
    \;\longrightarrow\; F(s) = d 
    \;\longrightarrow\; \phi(d).
\end{equation}
Here, $\mu_{S}$ denotes the scattering statistics that parametrize the generative maximum entropy model $p_{\mu_{S}}$(s), $s$ a realization sampled from it, $d = F(s)$ the map obtained by applying the forward operator in pixel space to $s$, and $\phi(d)$ its scattering statistics. 
The stochastic $\mathcal{F}$ forward operator then defines the likelihood function $p(\phi(d) \mid \mu_{S})$ in the new ($\mu_s$, $\phi(d)$) parametrization of the problem. 
However, this likelihood is analytically intractable since $\mathcal{F}$ involves both sampling from a gradient-descent generative model and applying the forward operator $F$ in pixel space. We therefore need to computationally approximate it.

One possible approach is to adapt simulation-based inference (SBI) to our problem~\citep[see, \textit{e.g.,}][]{Cranmer2020}. This would correspond to drawing samples $\mu_{S}$ from a proposal distribution $q(\mu_{S})$ (which is typically chose to be the prior distribution), and generating the corresponding $\phi(d)$ through the ST forward operator $\mathcal{F}$ in Eq.~\eqref{eq:effective forward}. These $(\mu_{S}, \phi(d))$ pairs would then be used to train a neural density estimator for the likelihood $p(\phi(d)\mid\mu_{S})$ \citep{papamakarios19}. In practice, however, training a neural density estimator in unstructured spaces of such high dimension, a few hundreds at least for both $\mu_S$ and $\phi(d)$, is challenging. 

To obtain a tractable approximation, we instead introduce an approximation for the ST forward operator $\mathcal{F}$ given Eq.~\ref{eq:effective forward} and thus for the likelihood. By performing a Taylor expansion of  $\mathcal{F}$ around any reference point $\mu_{S}$, we approximate the likelihood with a Gaussian linear model (see appendix \ref{app:Hypothesis on the likelihood function}):
\begin{equation}
    p(\phi(d) \mid \mu_{S}) \sim \mathcal{N} (A\mu_{S} +b \mid \Sigma). 
\end{equation}
Assuming such a linear Gaussian likelihood, as well as a uniform prior on $\mu_{S}$, the posterior is fully determined by Bayes rule and is also Gaussian.
Approximating the likelihood and the posterior then boils down to estimating $A$, $b$, and $\Sigma$, which can still be done in a SBI setting as described above.
We derive below an adaptive iterative algorithm adapted to the specificity of the ST space in order to approximate this posterior distribution.

\vspace{-0.3cm}
\subsection{Algorithm}\label{subsec:Algorithm}
\label{subsec:algorithm}

\begin{figure}[t!]
\begin{center}
    \includegraphics[scale=0.35]{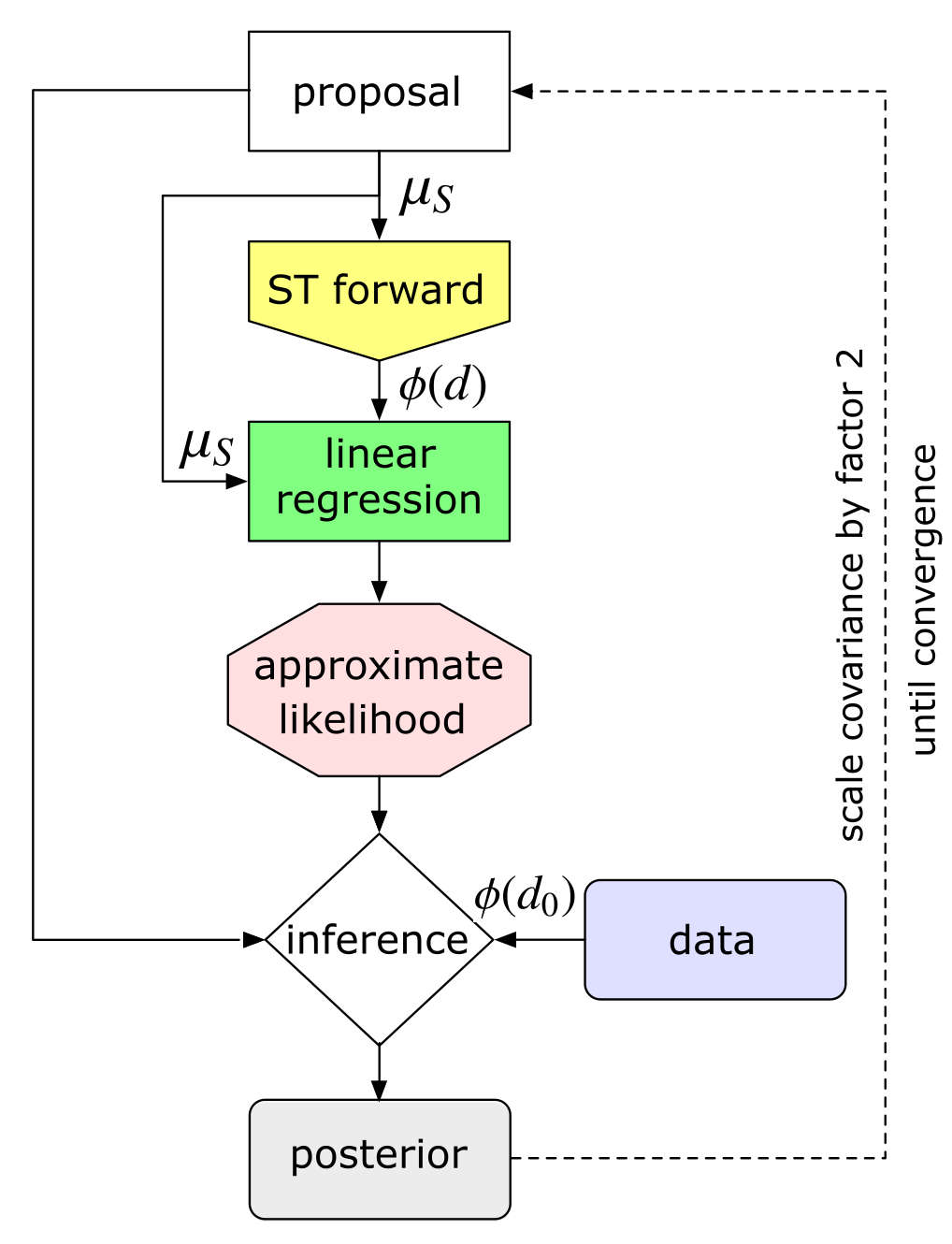}
    \caption{Schematic of the iterative algorithm used in this paper. See Sec.~\ref{subsec:algorithm} for more details.
    \vspace{-0.5cm}}
    \label{fig:schematic_pipeline}
\end{center}
\end{figure}

Our goal is to approximate the posterior distribution
\begin{equation}
p(\mu_{S} \mid \phi(d)) \propto p(\mu_{S})\mathcal{N}(A\mu_{S}+b \mid \Sigma),
\end{equation}
through the estimation of the parameters $A,b$ and $\Sigma$. To do so, we use an algorithm inspired by sequential likelihood estimation, in which an initial proposal distribution is iteratively refined to approximate the likelihood in increasingly relevant regions of the parameter space \citep{papamakarios19}.  A schematic of the algorithm, in the same style as \cite{Cranmer2020} is shown in Fig.\ref{fig:schematic_pipeline}. For brevity, the main text only provides its high-level description, and we refer to Appendix~\ref{app:convergence algorithm} for the mathematical details and convergence properties.

Starting from an initial proposal distribution $q_{0}(\mu_{S})$, we iteratively refine a sequence of proposals $q_{i}(\mu_{S})$. At each iteration~$i$, samples $\mu_{S} \sim q_{i}(\mu_{S})$ are propagated through the ST forward operator to generate $N_{pairs}$ pairs $\{\mu_{S}^{j}, \phi(d)^{j}\}$, which are then used to estimate $A_{i},b_{i}$ and $\Sigma_{i}$, and thus the likelihood, at step $i$. 

\textit{Initialization.} The initial proposal $q_{0}$ is chosen as a narrow Gaussian distribution whose mean is set to the ST statistics of a map $\tilde{s}$ that is in the region of high probability of the posterior. This reference map is obtained through a pixel-space optimization: starting from white noise, we optimize $\tilde{s}$ so as to minimize $|\phi(F(\tilde{s})) - \phi(d_{0})|^{2}$. By construction, the mean of $q_{0}$ therefore matches the observed statistics $\phi(d_{0})$ after application of the ST forward operator. This strategy has been shown to recover high-quality uncontaminated maps in~\cite{Regaldo2021} and was further adapted in~\cite{Delouis2022,Auclair2024}.

\textit{Sequential estimation.} Proposals are iteratively updated from the product between the last proposal and the estimated likelihood, which is Gaussian, and whose covariance is then scaled by $2$. At each step $N$, the resulting proposal $q_{N}(\mu_{S})$ is Gaussian with mean $\rho_{N}$ and covariance $\Gamma_{N}$ given by: 
\begin{equation}
\left\{
\begin{array}{l}
   \vspace{0.2cm} \Gamma_{N} = (\frac{1}{2^{N}}\Gamma_{0}^{-1} + \sum_{i=0}^{N-1} \frac{1}{2^{N-i}}A_{i}^{T}\Sigma_{i}^{-1}A_{i})^{-1}, \\
 \rho_{N} = \Gamma_{N}\cdot\left(\frac{1}{2^{N}}\Gamma_{0}^{-1}\rho_{0} + \sum_{i=0}^{N-1} \frac{1}{2^{N-i}}A_{i}^{T}\Sigma_{i}^{-1}(\phi(d_{0}) - b_{i})\right). 
\end{array}
\right.
\end{equation}
Scaling by $2$ the covariance between each proposal is needed to ensure that we do not introduce artificial overconfidence by conditioning multiple times on the same observation $\phi(d_{0})$. Indeed, at any step $N$, the relative weights assigned to the initial proposal and to the successive likelihood estimates satisfy: $\frac{1}{2^{N}}+ \sum_{i=0}^{N-1} \frac{1}{2^{N-i}} = 1$.

\textit{Stopping.} The iterations are stopped when a fixed point is reached, that is, when $q_{f}(\mu_{S}) = q_{f+1}(\mu_{S})$. At this iteration, the proposal coincides with the posterior for a uniform prior and a linear Gaussian likelihood, as shown in App.~\ref{app:convergence algorithm}.

\begin{figure*}[t!]
\includegraphics[scale=0.28]{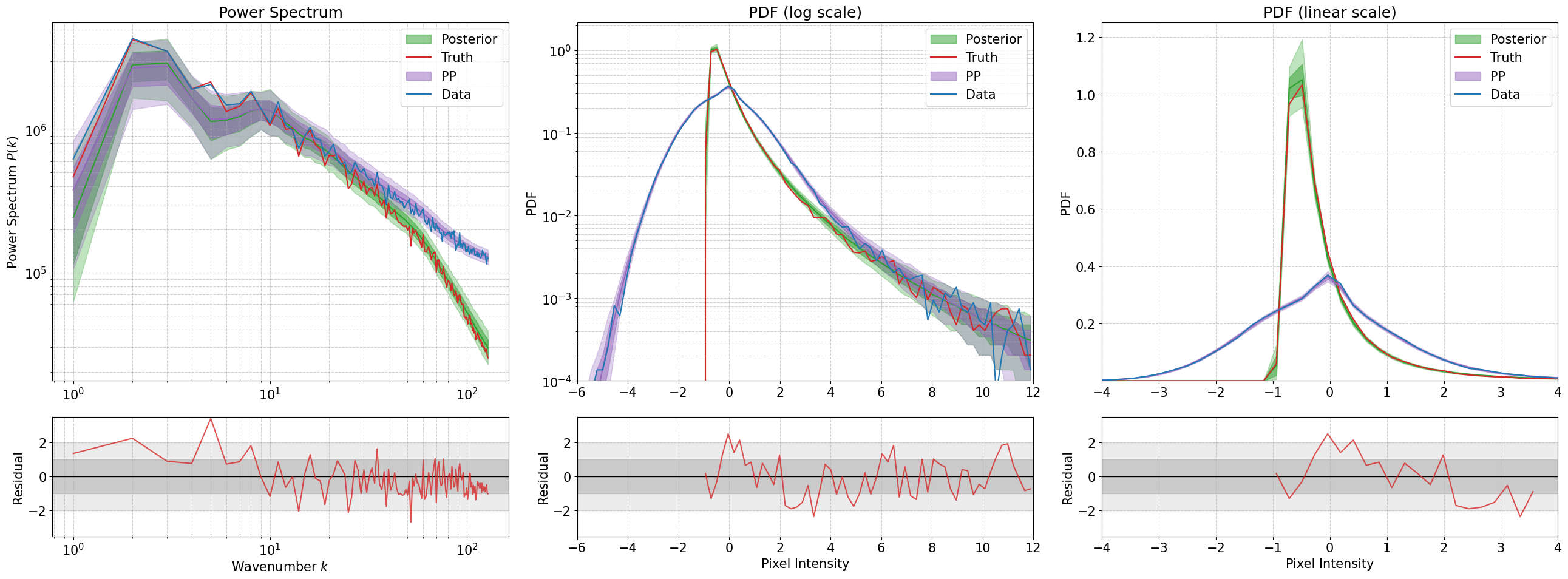}
\caption{\textbf{Top:} Comparison of summary statistics for the true field $s_{0}$ (red), the observation $d_{0}$ (blue), samples generated from the posterior distribution $p(\mu_{S} \mid \phi(d_{0}))$ (green), and the posterior predictive (PP) distribution (purple). The statistics are power spectrum, and one-point probability distribution function (PDF) shown in both linear and logarithmic scales. \textbf{Bottom:} Residual of the statistics of $s_{0}$ normalized by the standard deviation of the posterior distribution. 
We see that the posterior samples accurately reproduce the statistical properties of the true field, while the posterior predictive samples match those of the observation, up to sample variance.}

\label{fig:Astrophysical statistics}
\end{figure*}

\vspace{-0.3cm}
\section{Results}\label{sec:results}

\subsection{Data and setup}\label{subsec:data and setup}

We now provide a proof of concept of the methodology presented above for a specific field and forward operator $F$,
illustrating the capabilities and limitations of the approach.

\textit{Data}. 
We use as signal of interest $s_{0}$ a two-dimensional column density map extracted from a single realization of the \textit{Quijote} simulation suite \cite{VillaescusaNavarro2020}. 
This suit comprises a large set of cosmological $N$-body simulations that model the nonlinear gravitational evolution of dark matter from Gaussian initial conditions. 
We consider the same integrated
column density maps than those used in~\cite{Allys2020}, projected into $256\times256$ maps for a region of 1000$\times$1000 ($h^{-1}\rm{Mpc})^2$ at $z=0$. These density fields exhibit strong non-Gaussian features and structures across a wide range of spatial scales.

\textit{Forward operator}. 
We consider a forward operator designed to mimic instrumental contamination\footnote{The goal is to capture generic instrumental effects. This forward operator is not meant to be fully realistic, as large-scale structure fields are not directly observed directly through an instrument in practice.}. It consists of adding white Gaussian noise to the maps, followed by the application of elliptical masks. The number, positions, and sizes of the masks are drawn randomly at each application of the forward operator as described in App.~\ref{app:numerical_details}. Since our goal is to perform a statistical reconstruction of $s_{0}$ rather than a deterministic one, this randomness is not a limitation. Overall, the forward operator is linear, stochastic, and non-invertible. The amplitude of the Gaussian noise is chosen such that the signal-to-noise ratio of the resulting contaminated map is equal to $1$.

\textit{ST generative model}. To build our ST generative model, we use the scattering spectra statistics introduced in~\cite{Morel2023} and~\cite{Cheng2024}. These statistics characterize the power and sparsity at given oriented scales as well as the interaction between different oriented scales. Their detailed description is provided in Appendix~\ref{app:ScatteringStatistics}. We also describe in this appendix how we compress this set of ST statistics, following an approach firstly introduced in~\cite{Allys2019_RWST} and generalized in~\cite{Cheng2024}. This reduction allows us to build generative models with a moderate number of parameters, $l = 243$, which is critical for our approach.

For \textit{Quijote} maps, we also construct an ST model of the logarithm of the density field, instead of the density field itself. We then take the exponential of this logarithmic field to recover a realization of the raw density field. Indeed, it has been shown in~
\citep{Allys2020,Mousset2024}  that this approach is more effective to build a good ST model of the density field, both statistically and visually. This approach is also motivated by the strict positivity of the density field. We provide all additional numerical details of our experiments in App.~\ref{app:numerical_details}.

\begin{figure*}[t!]
\includegraphics[scale=0.45]{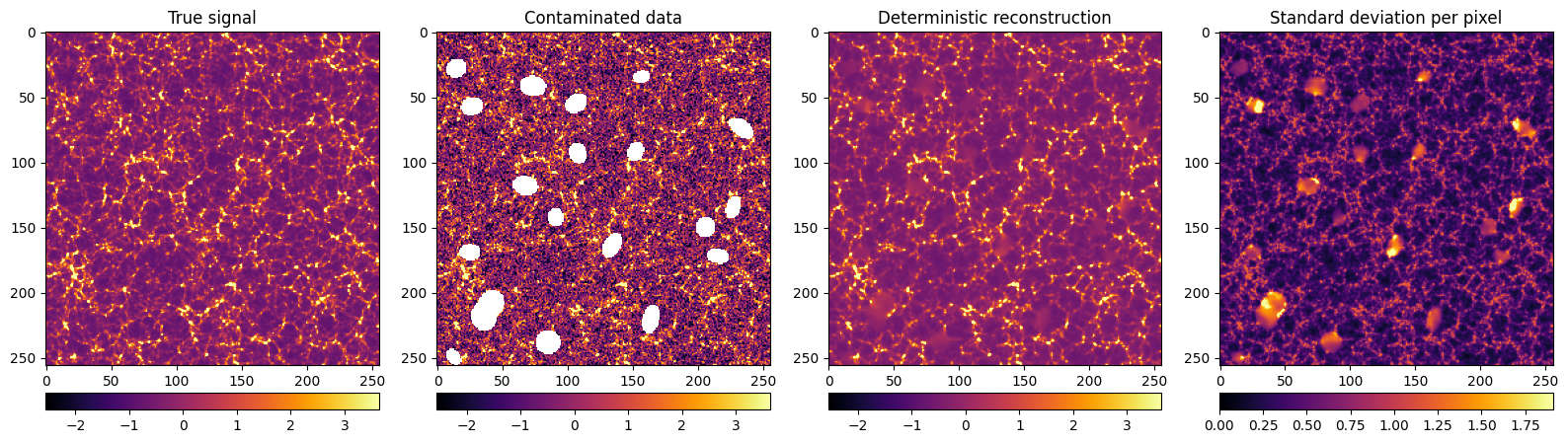}
\caption{True signal $s_{0}$, observation $d_{0}$, and posterior mean and standard deviation maps, both estimated pixel-wise using a moment network. The posterior mean recovers the large-scale features of the true signal while filtering the noise-dominated small scales. In the masked regions, where the data provide no information, the posterior uncertainty is large, as expected in a Bayesian framework. }
\label{fig:determinstic reconstruction}
\end{figure*}

\vspace{-0.2cm}

\subsection{Visual validation}\label{subsec:Visual validation}

After inferring the posterior distribution on ST statistics, $p(\mu_{S} \mid \phi(d_{0}))$, we generate maps by sampling the ST generative model conditioned on $\mu_s$ drawn from this posterior. Note that these samples are not expected to deterministically match the pixel-level structure of the true signal $s_{0}$.
From these samples in pixel space, we can construct approximate posterior distributions for any summary statistics, as well as posterior predictive distributions by applying the forward operator $F$ to them.

The first line of Fig.~\ref{fig:visual samples} shows $s_{0}$ and three samples of the posterior. To check the quality of these samples, we also show corresponding predictive samples as well as $d_{0}$ in the bottom row. We recall that our goal is to find a distribution of ST models, as wide as possible, such that the generated maps from these models could not be distinguishable from the observation after applying the forward operator. These three predictive samples are visually indistinguishable from the data. In addition, the corresponding posterior samples appear very consistent with the \textit{Quijote} field, with variability mainly at the smallest, noise-dominated scales.
\vspace{-0.2cm}
\subsection{Statistical validation}\label{subsec:Statistical validation}

While visual agreement is already a difficult and important validation, we also check whether our posterior distribution reproduces standard cosmological statistics. In Fig.~\ref{fig:Astrophysical statistics}, we compare the power spectrum, the Probability Distribution Function (PDF) for the true signal  $s_{0}$ (red), our posterior distribution (green), the data $d_{0}$ (blue), and the posterior predictive distribution (purple). We also present three Minkowski functionals, which are statistics aiming at capturing the topology of the fields \citep{Schmalzing1996}, in Fig.~\ref{fig:Minkowski statistic} in Appendix.~\ref{app:Additional figures}.

The posterior distribution recovers very well all statistics of $s_{0}$. The posterior predictive distribution systematically matches the statistics of $d_{0}$, showing that none of these different summary statistics can distinguish the observation from our posterior predictive samples. It demonstrates that our method identifies a distribution of ST statistics whose generated maps are statistically consistent with the data under the given forward operator.
\vspace{-0.2cm}
\subsection{Pixel-level reconstruction}\label{subsec:Pixel-level reconstruction}

Our main objective in this work was to infer a distribution of models $p(\mu_{S} \mid \phi(d_{0}))$, parametrized by ST statistics, whose generated maps are statistically compatible with the observation once the forward operator is applied. As a consequence, the samples obtained from this distribution are only required to be statistically consistent with the signal of interest, rather than deterministically trying to reproduce its pixel-level structure.

If a deterministic pixel-level reconstruction is desired, it can be performed as a subsequent step after inferring this distribution of models. We follow here exactly the approach introduced in~\cite{Jeffrey2022}, to which we refer for more details. Specifically, we generate samples in pixel space from the inferred ST posterior, apply the forward operator to obtain the corresponding contaminated maps, and then train a neural network to remove the contamination. An example of such a reconstruction is shown in Fig.~\ref{fig:determinstic reconstruction}, where a moment network is used to estimate, for each pixel, both the posterior mean and standard deviation. The maps reconstructed from the posterior mean reproduce the large scale structure of the true signal deterministically, while naturally smoothing the small scale features that are most affected by contamination. Inside the masked regions, where the data provide no information, the posterior standard deviation becomes large, as expected, indicating that a wide range of pixel values is compatible with the observation.

\vspace{-0.3cm}
\section{Conclusion}\label{sec:conclusion}

In this work, we introduce a framework to solve astrophysical/cosmological imaging inverse problems in a challenging setting. We are in a low-data regime with a single (or very few) observation $d_{0}$, no external prior model for the signal of interest $s_{0}$, a stochastic and non invertible but known forward operator $F$. Using the possibility of constructing efficient maximum-entropy generative models of physical fields based on their ST statistics, we propose to shift the inference objective from a pixel-space Bayesian reconstruction of $s_{0}$ to the recovery of its ST statistics.   

Within this framework, we approximated the likelihood $p(\phi(d) \mid \mu_{S})$ in ST space by a linear Gaussian model, naturally motivated by a Taylor expansion of the ST forward operator. This approximation enables a tractable Bayesian treatment while remaining sufficiently expressive to capture the effect of the forward operator in ST statistics space. To estimate this likelihood in practice, we proposed an iterative algorithm inspired by sequential likelihood estimation, in which an adaptive proposal distribution is progressively refined to focus on regions of high posterior probability. 

Applied to \textit{Quijote} large-scale structure simulations and a stochastic, non-invertible instrumental forward operator, our method successfully recovers a posterior distribution over ST statistics. Samples drawn from this posterior generate fields that are statistically and visually consistent with the signal $s_{0}$, and that are indistinguishable from the data $d_{0}$ after applying the forward operator. This demonstrates that the inferred ST posterior can explain the observation under the forward operator, despite the absence of an external physical prior on the signal.

While the present study is limited to a simplified setup, it opens several promising avenues for future work. In particular, this approach is well suited to non-Gaussian signals that are difficult to simulate accurately, such as Galactic dust emission, where simulation-based priors are either unavailable or unreliable. Extensions of this framework could include the incorporation of additional data (\textit{e.g.}, multiple frequencies or ancillary observations) or the introduction of complementary constraints in ST space. More broadly, this study highlights the potential of ST representations as an alternative inference space for complex inverse problems in astrophysics and cosmology.

\begin{acknowledgements}
We sincerely thank S. Clark, J.-M. Delouis, I. Grenier, S. Mallat, and B. Wandelt for their valuable feedback. This work received government funding managed by the French National Research Agency under France 2030, reference numbers “ANR-23-IACL-0008” and "ANR-25-CE46-6634", as well as from the Paris Region under the DIM ORIGINE funding.
This work was granted access to the HPC resources of MesoPSL financed by the Region Ile de France and the project Equip@Meso (reference ANR-10-EQPX-29-01) of the programme Investissements d’Avenir.
\end{acknowledgements}

\vspace{-0.3cm}

\bibliographystyle{bibtex/aa.bst}

\bibliography{bib}

\begin{appendix}\label{app:appendices}

\section{Additional figures}
\label{app:Additional figures}

\begin{figure*}[t!]
\includegraphics[scale=0.40]{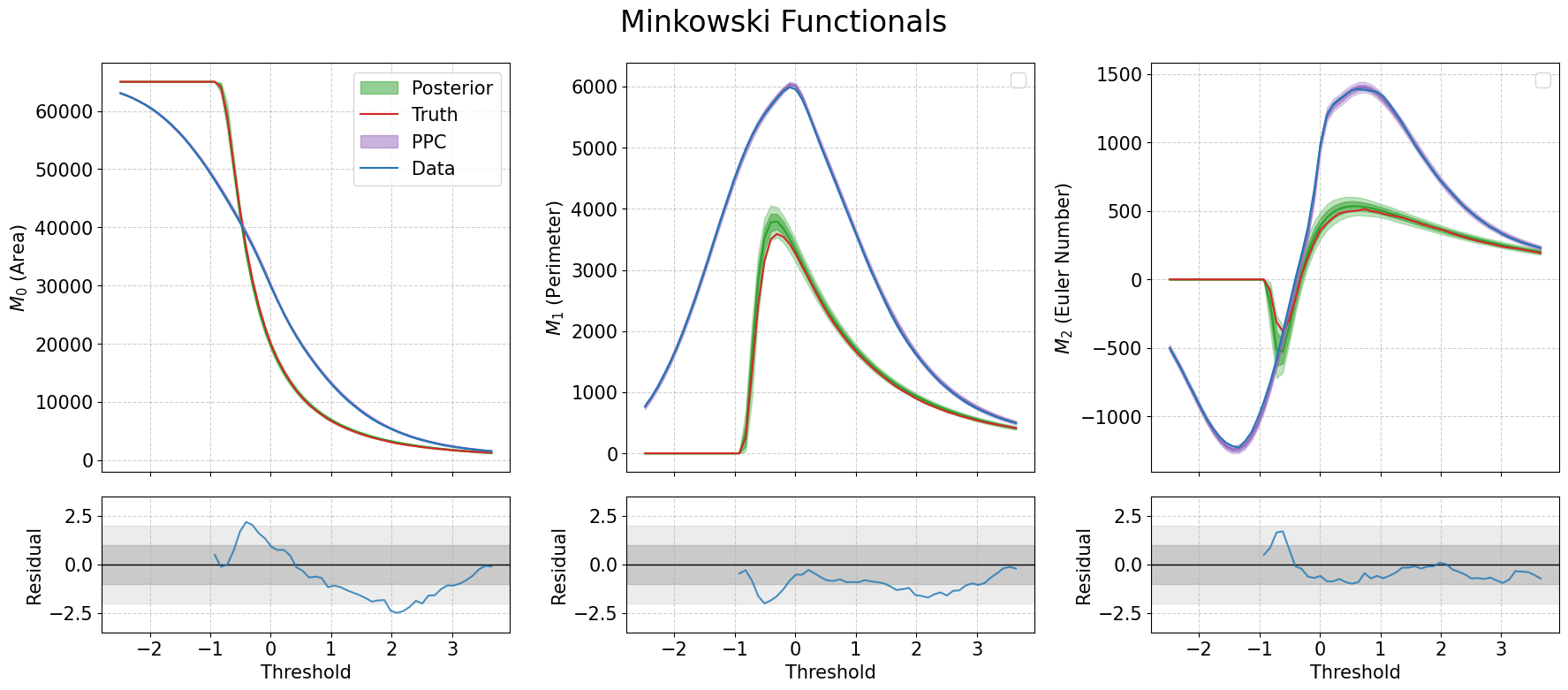}
\caption{Comparison of Minkowski functionals for the true field $s_{0}$ (red), the observation $d_{0}$ (blue), samples generated from the posterior distribution $p(\mu_{S} \mid \phi(d_{0}))$ (green), and the posterior predictive distribution (purple). 
We see that the posterior samples accurately reproduce the statistical properties of the true field with a slight bias for $M1$, while the posterior predictive samples match those of the observation, demonstrating the consistency of the inferred posterior with the data under the forward model.}
\label{fig:Minkowski statistic}
\end{figure*}

Fig.~\ref{fig:Minkowski statistic} complements Fig.~\ref{fig:Astrophysical statistics} with the estimated Minkowski functionals of the different fields (true, observed, sampled from the posterior, and for the posterior predictive distribution). 

\section{Hypothesis on the likelihood function}
\label{app:Hypothesis on the likelihood function}

Accounting for all stochasticity in the ST forward operator (stochasticity form the pixel space forward operator, intrinsic stochasticity of the generative model,...) with a variable $\xi$ we have
\begin{equation}
    \phi(d) = \mathcal{F}(\mu_{S}, \xi).
\end{equation} 
The likelihood is then given by marginalising over the stochasticity $\xi$ such that: 
\begin{equation}
    p(\phi(d) \mid \mu_{S}) = \int \delta\!\big(\phi(d) - \mathcal{F}(\mu_{S}, \xi)\big)\, p(\xi)\, d\xi.
\end{equation}
As $\mathcal{F}$ is non linear in both $\mu_{S}$ and $\xi$, this likelihood has no closed form. By performing a Taylor expansion around any point $\mu_{S}^{0}$ we can linearize the mean behavior of $\mathcal{F}$: 
\begin{equation}
    \mathbb{E}_{\xi} [ \mathcal{F}(\mu_{S},\xi) ] \approx \mathbb{E}_{\xi}[\mathcal{F}(\mu_{S}^{0},\xi)] + J_{\mu_{S}}(\mu_{S} - \mu_{S}^{0}).
\end{equation}
The Jacobian $J_{\mu_{S}}$ is the local sensitivity of the stochastic forward operator $\mathcal{F}$ with respect to its input parameters, evaluated at the reference point $\mu_{S}^{0}$: $J_{\mu_{S}} = \mathbb{E}_{\xi} \!\left[ 
    \frac{\partial \,\mathcal{F}(\mu_{s}, \xi)}{\partial \mu_{s}} 
\right] |_{\mu_{s}^{(0)}}.$
For a given $\mu_{S}$ and $\xi$ we have: 
\begin{equation}
    \mathcal{F}(\mu_{S},\xi) = \mathbb{E}_{\xi} [ \mathcal{F}(\mu_{S},\xi) ] + \epsilon(\mu_{S}, \xi).
\end{equation}
The last term $\epsilon(\mu_{S}, \xi)$ represents the random deviation of a sample from the mean prediction at that parameter value. It has zero mean by definition and we will assume that it is Gaussian distributed. It's covariance $\Sigma(\mu_{S})$ is also the covariance induced by the stochasticity of the ST forward operator: 
\begin{equation}
    \Sigma(\mu_{S}) = \mathbb{E}_{\xi}(\epsilon(\mu_{S},\xi)\epsilon(\mu_{S},\xi)^{T}) = Cov_{\xi}(\mathcal{F}(\mu_{s},\xi)).
\end{equation}
We will also make the hypothesis that in the vicinity of the expansion point $\mu_{S}^{0}$ this covariance term does not depend on $\mu_{S}$. Using our different hypothesis, we get:
\begin{equation}
    \phi(d) \approx \mathbb{E}_{\xi}[\mathcal{F}(\mu_{S}^{0},\xi)] + J_{\mu_{S}}(\mu_{S} - \mu_{S}^{0}) + \epsilon, \ \ \epsilon \sim \mathcal{N}(0,\Sigma),
\end{equation}
which corresponds to a linear Gaussian model for the likelihood:
\begin{equation}
    p(\phi(d) \mid \mu_{S}) \sim \mathcal{N} (A\mu_{S} +b \mid \Sigma). 
\end{equation}

\section{Set of scattering statistics used}
\label{app:ScatteringStatistics}

In this work, the set of ST statistics we use are scattering covariances, introduced in~\cite{Morel2023,Cheng2024}, and we refer to these papers for a more comprehensive introduction. Scattering covariance statistics are constructed through successive wavelet transforms and modulus operators applied to a field $I$, followed by spatial mean or covariance estimations. Each oriented wavelet $\psi^{\lambda_{i}}$ is indexed by $\lambda_{i} = (j_{i}, \theta_{i})$, denoting its dyadic scale as well as orientation. We used $7$ dyadic scales and $4$ orientations between 0 and $\pi$. We consider 4 types of coefficients: 
\vspace{-0.3cm}
\begin{equation}\label{eq:S1}
    S_{1}^{\lambda_{1}} = \langle | I\star \psi^{\lambda_{1}} | \rangle,
\end{equation}
\vspace{-0.6cm}
\begin{equation}\label{eq:S2}
    S_{2}^{\lambda_{1}} = \langle | I\star \psi^{\lambda_{1}} |^{2} \rangle,
\end{equation}
\vspace{-0.6cm}
\begin{equation}\label{eq:S3}
    S_{3}^{\lambda_{1},\lambda_{2}} = Cov \left[I \star \psi^{\lambda_{1}}, |I \star \psi^{\lambda_{2}}| \star \psi^{\lambda_{1}} \right],
\end{equation}
\vspace{-0.6cm}
\begin{equation}\label{eq:S4}
    S_{4}^{\lambda_{1},\lambda_{2},\lambda_{3}} =  Cov \left[|I \star \psi^{\lambda_{3}}| \star \psi^{\lambda_{1}}, |I \star \psi^{\lambda_{2}}| \star \psi^{\lambda_{1}} \right],
\end{equation}
where $\langle \cdot \rangle$ is a spatial average and $Cov \left[XY\right] = \langle XY^{*} \rangle - \langle X \rangle \langle Y^{*} \rangle$ is the estimated spatial covariance for two complex fields $X$ and $Y$. As $|I \star \psi^\lambda|$ only contains spatial frequencies at larger scales than $\psi^\lambda$, as well as to avoid duplicate terms, only $\lambda$ configurations that verify $j_3 \leq j_2 \leq j_1$ are considered. For maps of size $256\times256$, the number of coefficients is then around $9000$. These coefficients are all normalized like in \cite{Cheng2024} with a reference map that we choose as the mean of the first proposal $\phi(\tilde{s})$.

As our approach relies on computing linear regression and estimating covariances, reducing this dimension is critical. The first dimensional reduction that we perform, which is valid for isotropic fields, is to reduce the angular dependence of ST statistics to only relative angles $\theta_{2}-\theta_{1}$ and $\theta_{3}-\theta_{1}$. We therefore reparametrize the ST statistics accordingly and average over $\theta_{1}$ to take isotropy into account. The second dimensional reduction that we perform exploits the strong regularity usually exhibited by ST statistics of physical fields as function of scales $j$ and angles $\theta$. To do so, we follow the approach introduced in \citep{Allys2019_RWST} and generalized in \citep{Cheng2024}. For the angular regularity, we perform a Fourier transform with respect to the re-parametrized variables $\theta_{2}-\theta_{1}$ and $\theta_{3}-\theta_{1}$, and then keep only the zeroth mode and the first two harmonics. For the scale regularity, the ST statistics are first re-parametrized using a reference scale $j_{1}$ and scale differences $j_{2}-j_{1}$ and $j_{3}-j_{1}$. Then, a cosine transform is performed along $j_1$ only, and only the zeroth mode and the first two harmonics are also retained. This projection\footnote{The number of retained harmonics can be adjusted depending on the process studied.} yields a final compressed representation of dimension $243$.


\section{Mathematical details of the algorithm}
\label{app:convergence algorithm}

We give here the mathematical details of the algorithm described in
Sect.~\ref{subsec:Algorithm}. The key identity used throughout is that the
product of a Gaussian proposal
$q(\mu_{S}) \sim \mathcal{N}(\rho, \Gamma)$ and a linear Gaussian likelihood
with parameter-independent covariance,
$p(\phi(d)\mid \mu_{S}) \sim \mathcal{N}(A\mu_{S} + b, \Sigma)$,
is itself Gaussian and admits a closed-form expression.

The resulting distribution has mean $m$ and covariance $C$ given by
\begin{equation}
\left\{
\begin{array}{l}
   \vspace{0.2cm}
   C = \left(\Gamma^{-1} + A^{\top}\Sigma^{-1}A\right)^{-1}, \\
   m = C\left(\Gamma^{-1}\rho + A^{\top}\Sigma^{-1}(\phi(d_{0}) - b)\right).
\end{array}
\right.
\end{equation}

Using the proposal update rule introduced in Sect.~\ref{subsec:Algorithm},
this identity yields a recurrence relation for the mean and covariance
$(\rho_{N}, \Gamma_{N})$ of the successive proposal distributions
$q_{N}(\mu_{S})$:
\begin{equation}
\left\{
\begin{array}{l}
   \vspace{0.2cm}
   \Gamma_{N+1}
   = 2\left(\Gamma_{N}^{-1} + A_{N}^{\top}\Sigma_{N}^{-1}A_{N}\right)^{-1}, \\
   2\,\Gamma_{N+1}^{-1}\rho_{N+1}
   = \Gamma_{N}^{-1}\rho_{N}
     + A_{N}^{\top}\Sigma_{N}^{-1}\left(\phi(d_{0}) - b_{N}\right).
\end{array}
\right.
\end{equation}
The factor $2$ appearing in both equations arises from the rescaling of the
proposal covariance by a factor of $2$ at each iteration. This rescaling is
introduced to avoid over-conditioning on the same observation $\phi(d_{0})$,
which would otherwise lead to artificially overconfident proposals when the
estimated likelihood is repeatedly incorporated. Solving this recurrence relation give the formula presented in \ref{subsec:Algorithm} of the mean and covariance of the proposal at any step $N$: 
\begin{equation}
\left\{
\begin{array}{l}
   \vspace{0.2cm} \Gamma_{N} = (\frac{1}{2^{N}}\Gamma_{0}^{-1} + \sum_{i=0}^{N-1} \frac{1}{2^{N-i}}A_{i}^{T}\Sigma_{i}^{-1}A_{i})^{-1}, \\
 \rho_{N} = \Gamma_{N}\cdot\left(\frac{1}{2^{N}}\Gamma_{0}^{-1}\rho_{0} + \sum_{i=0}^{N-1} \frac{1}{2^{N-i}}A_{i}^{T}\Sigma_{i}^{-1}(\phi(d_{0}) - b_{i})\right). 
\end{array}
\right.
\end{equation}
This expression shows that the influence of the initial proposal is
exponentially suppressed as $N$ increases, while all previously estimated
likelihoods contribute to the proposal update with exponentially decreasing
weights. We define convergence as reaching a fixed point of the iterative process, that is, when two successive proposals are identical. While we do not provide a theoretical proof of convergence, this behavior is consistently observed in practice. Convergence is assessed by monitoring the Kullback–Leibler (KL) divergence between successive Gaussian proposals, which admits a closed-form expression and is therefore inexpensive to evaluate. In practice, this divergence never reaches zero due to the finite sampling of each proposal, which induces noise in the estimation of $A$, $b$, and $\Sigma$. We therefore stop the algorithm when the KL divergence reaches a plateau. At the convergence step $f$, we have:

\begin{equation}
\left\{
\begin{array}{l}
   \vspace{0.2cm}
   \Gamma_{f+1}
   = 2\left(\Gamma_{f}^{-1} + A_{f}^{\top}\Sigma_{f}^{-1}A_{f}\right)^{-1} = \Gamma_{f}, \\
   2\,\Gamma_{f+1}^{-1}\rho_{f+1}
   = \Gamma_{f}^{-1}\rho_{f}
     + A_{f}^{\top}\Sigma_{f}^{-1}\left(\phi(d_{0}) - b_{f}\right) = 2\Gamma_{f}^{-1}\rho_{f}.
\end{array}
\right.
\end{equation}
Which reduces to: 
\begin{equation}
\left\{
\begin{array}{l}
   \vspace{0.2cm}
   \Gamma_{f}
   = \left(A_{f}^{\top}\Sigma_{f}^{-1}A_{f}\right)^{-1}, \\
   \rho_{f}
   = \Gamma_{f}A_{f}^{\top}\Sigma_{f}^{-1}\left(\phi(d_{0}) - b_{f}\right).
\end{array}
\right.
\end{equation}
This corresponds to the posterior distribution for a uniform prior and a Gaussian linear model for the likelihood estimated from the $q_{f}(\mu_{S})$ proposal distribution.


\section{Additional numerical details}
\label{app:numerical_details}

At each application of the forward operator, the number, position, and shape of the masks are randomly drawn. The number of masks is sampled uniformly between $10$ and $20$, and their positions are drawn uniformly over the map. The masks have elliptical shapes, with semi-axes sampled from a lognormal distribution of mean $2$ and variance $0.25$, and random orientations drawn uniformly in $[0,2\pi]$. While these choices are arbitrary, they are designed to mimic typical instrumental masking patterns.

At each iteration, the parameters of the Gaussian likelihood $p(\phi(d)\mid\mu_{S})$ are estimated from $4000$ samples drawn from the current proposal distribution. The ST generative model is sampled by batches of $200$ on an \textit{A100 80GB NVIDIA GPU}. Within each epoch, the first batch is generated starting from white Gaussian noise, while subsequent batches are initialized from the output of the previous batch. This warm-start strategy significantly accelerates the overall algorithm.
The linear regression used to estimate $A$
and $b$ is performed using ridge regression. The corresponding
regularization parameter is selected by cross-validation on the proposal
samples.

The initial proposal covariance is set to $\Gamma_{0} = 10^{-4} \cdot \mathrm{Id}$.
Although convergence is observed after about 15 epochs, the algorithm is run for 40 epochs in order to be conservative. After epoch 15, the proposals change only marginally.
A small diagonal regularization term of amplitude $10^{-8}$ is systematically
added when estimating and inverting covariance matrices to ensure
numerical stability and well-conditioned inverses.

Although the first likelihood estimates (and thus proposals) obtained at early iterations may be noisy due to the limited number of samples, the iterative nature of the algorithm progressively refine the estimates of $A,b$ and $\Sigma$. As a result their estimation stabilizes as the algorithm converges.

\end{appendix}
\end{document}